\documentclass[%
 aip,
 amsmath,amssymb,
 preprint,%
]{revtex4-1}

\usepackage{graphicx}
\usepackage{dcolumn}
\usepackage{bm}

\usepackage[utf8]{inputenc}
\usepackage[T1]{fontenc}
\usepackage{mathptmx}
\usepackage{etoolbox}

\usepackage{color}
\usepackage{physics}
\usepackage{mathtools}
\usepackage{soul}

\makeatletter
\def\@email#1#2{%
 \endgroup
 \patchcmd{\titleblock@produce}
  {\frontmatter@RRAPformat}
  {\frontmatter@RRAPformat{\produce@RRAP{*#1\href{mailto:#2}{#2}}}\frontmatter@RRAPformat}
  {}{}
}%
\makeatother
\begin{document}

\preprint{AIP/123-QED}

\title[How to grow an oscillator network with enhanced synchronization]
{How to grow an oscillators’ network with enhanced synchronization}
\author{Jong-Min Park}
 \affiliation{
 School of Physics, Korea Institute for Advanced Study, Seoul 02455, Republic of Korea
 }
 
\author{Daekyung Lee}%

\author{Heetae Kim}
\email{hkim@kentech.ac.kr}
\affiliation{%
Department of Energy Engineering, Korea Institute of Energy Technology, Naju 58330, Republic of Korea
}%

\date{\today}

\begin{abstract}
We study a way to set the natural frequency of a newly added oscillator in a growing network to enhance synchronization. 
Population growth is one of the typical features of many oscillator systems for which synchronization is required to perform their functions properly. 
Despite this significance, little has been known about synchronization in growing systems. We suggest effective growing schemes to enhance synchronization as the network grows under a predetermined rule. 
Specifically, we find that a method based on a link-wise order parameter outperforms that based on the conventional global order parameter. 
With simple solvable examples, we verify that the results coincide with intuitive expectations. 
The numerical results demonstrate that the approximate optimal values from the suggested method show a larger synchronization enhancement in comparison to other naive strategies. 
The results also show that our proposed approach outperforms others over a wide range of coupling strengths.
\end{abstract}

\maketitle

\begin{quotation}
Many dynamical systems show periodic motion that couples the affiliated elements on a complex interaction structure and perform various functions in a synchronized state.
Most of them have a common feature that their system size is not fixed but grows.
Thus we study the synchronization of the coupled oscillators in a growing system.
We find that one can enhance the synchronization by assigning a proper natural frequency to newly added dynamical nodes.
In addition, we suggest a useful prediction for the optimal natural frequency.
It helps us to understand how to evolve oscillator systems, keeping their functional stability.
\end{quotation}

\section{Introduction}

Many natural and artificial oscillator systems must be in a synchronized state to carry out their functions properly, including power-grid networks~\cite{power_grid}, heart cells, neural networks, biological circadian clocks~\cite{circadian}, etc.
For stable operation, maintaining the synchronized state is important for such networks even in the presence of newcomers.
However, few studies have been conducted on how an added oscillator affects the coherent behavior of the system~\cite{SBG}. In this context, it is worth studying the optimal condition for a newly added oscillator to enhance the synchronization of a growing network.

The Kuramoto model is one of the popular models that describe the synchronization phenomenon between coupled oscillators.
The original Kuramoto model describes a system of free oscillators at their natural frequency with all-to-all pairwise interaction between them~\cite{kuramoto2003chemical,acebron2005kuramoto}.
In many real systems, though, the interaction is inhomogeneous.
In this case, the system can be described as Kuramoto oscillators in a network, where the nodes and links correspond to the oscillators and the interactions, respectively.
One of the significant questions is how the topology of the interaction influences the degree of synchronization~\cite{huang2006abnormal,gomez2007paths,gomez2007synchronizability,arenas2008synchronization,gomez2011explosive,skardal2013effects}.

In attempts to answer this question, several studies have investigated the effect of the interaction topology on synchronization in various network structures such as small-world networks~\cite{niebur1991oscillator,hong2002synchronization,barahona2002synchronization}, scale-free networks~\cite{moreno2004synchronization,mcgraw2007analysis,hong2007finite}, regular lattices~\cite{strogatz1988collective,hong2004collective,hong2005collective,hong2007entrainment}, and motifs~\cite{vega2004fitness}.
These studies have shown that the topological properties of interactions significantly influence the nature of the synchronization phase transition.
To understand this correlation between network structure and dynamics analytically, various mean-field approaches have been developed~\cite{ichinomiya2004frequency,restrepo2005onset,ichinomiya2005path,lee2005synchronization,restrepo2006emergence}.

In particular, previous studies~\cite{skardal2014optimal,pinto2015optimal,Lei2022unified} have suggested ways to find the optimal structure or the natural frequency distribution by introducing objective functions obtained from macroscopic order parameters in the strong coupling limit.
They have shown that by searching the structure or frequency distribution that lowers the objective functions, one can approach the optimal condition.
However, these methods were only applied to systems with a fixed size.

In this study, 
we demonstrate that one can utilize the same approach as in Refs.~\onlinecite{skardal2014optimal,Lei2022unified} to find an effective condition for enhancing synchronization in growing systems. To be precise, we find the optimal natural frequency of a newly added oscillator as the network evolves though a predetermined process.
This paper is organized as follows.
In Sec.~\ref{sec:model}, we explain the model and our main goal, and in Sec.~\ref{sec:derivation} we present our main results.
The results are applied to simple tractable cases in Sec.~\ref{sec:simple_ex}
and are numerically verified in more complicated growing networks in Sec.~\ref{sec:num_verification}.
In Sec.~\ref{sec:conclusion}, we give the conclusions of this paper.

\section{Model}\label{sec:model}
\begin{figure}
\centering
\includegraphics[width = \columnwidth]{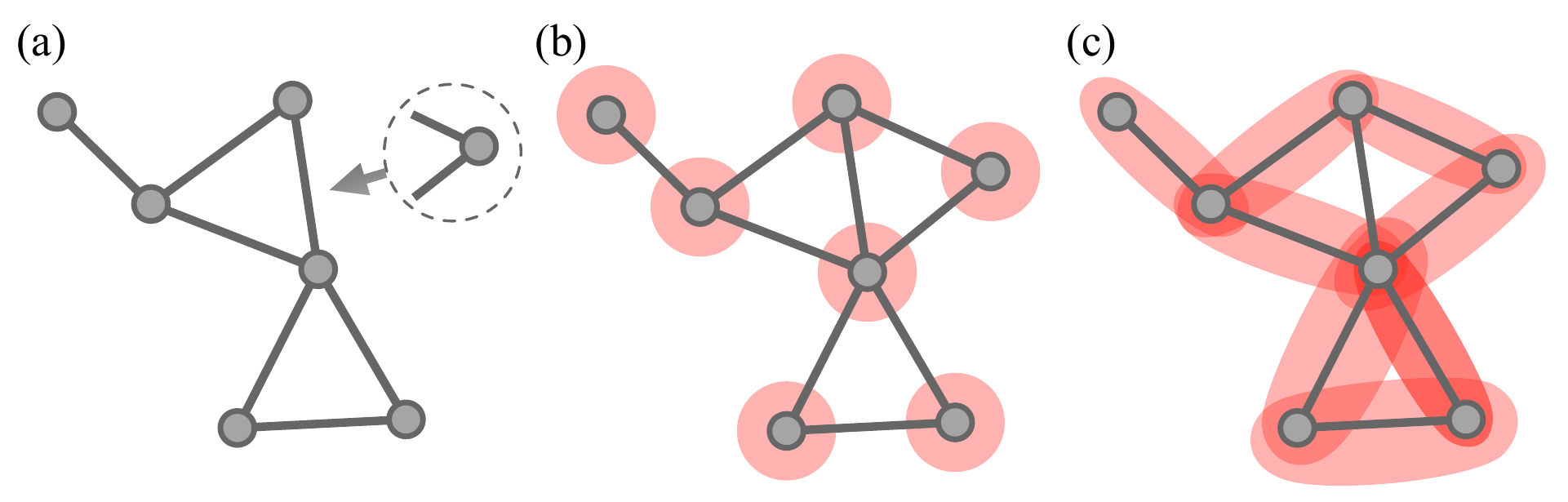}
\caption{Conceptual illustration of (a) a growing network, (b) synchrony alignment function (SAF), and (c) adjusted Lyapunov function (ALF). (a) In each step of a growing network, a new node is introduced to the existing network such that it participates in the synchronization dynamics. (b) As a strategy to assign the natural frequency of the new node, the SAF method searches the condition to maximize the system's order parameter, which only focuses on each node's phase. (c) In contrast, the ALF method tries to optimize the sum of the local order parameters, which consider the phase difference between each node pair.
}
\label{fig:scheme}
\end{figure}

We consider partially coupled Kuramoto oscillators. The interaction structure is revealed through the network topology, where dynamic nodes correspond to oscillators and links indicate pairs of the interacting nodes. To take into account system growth, we consider the interaction network to grow under a predetermined rule. Here, we suppose that only a single node is added at each growing step [see Fig.~\ref{fig:scheme} (a)]. In this setup, our purpose is to find the optimal natural frequency of the new node that leads to the highest degree of synchronization.

After a single growing step, the system's equation of motion is given by the Kuramoto equation~\cite{kuramoto2003chemical},
\begin{equation}\label{eq:eq_of_motion}
    \dot{\theta}_i(t) =
    \omega_i -
    K \sum_{j=0}^N a_{ij} \sin \left (\theta_i(t)-\theta_j(t) \right)~,
\end{equation}
where $\theta_i(t)$ and $\omega_i$ are the phase and the natural frequency of the $i$-th oscillator, respectively, for $i=0, 1, ..., N$ with the initial system size $N$, $K$ is the coupling strength, and $a_{ij}$ is the adjacency matrix characterizing the interaction network.
Here, the index $i=0$ is used to indicate the newly added node.

To clarify our purpose, we need to consider the macroscopic measure of the degree of synchronization, namely the Kuramoto order parameter defined as 
\begin{equation}\label{def:order_para}
    r =
    \left | \frac{1}{N+1} \sum_{j=0}^N e^{i \theta_j(t)} \right |~.
\end{equation}
With this, we express our main goal as finding the $\omega_0$ with which the order parameter $r$ takes its maximum.

In the next section, we find the optimal $\omega_0$ in the strong coupling regime $K \gg |\omega_i|$, which gives an approximate solution.
The previous study~\cite{Lei2022unified} showed that one may use alternative order parameters to yield better approximations.
Here, we adopt an additional macroscopic measure, referred to as the total local order parameter [see Fig.~\ref{fig:scheme} (c)], given by
\begin{equation}\label{def:local_order_para}
    \tilde{r} =
    \sum_{j=0}^N \tilde{r}_j \equiv
    \sum_{j,k} a_{jk} \cos \left( \theta_j(t) - \theta_k(t) \right)~.
\end{equation}
The local order parameter $\tilde{r}_i$ was originally introduced to understand the effect of the network topology on the formation of synchronized clusters in the relaxation process~\cite{arenas2006synchronization,arenas2006synchronization2,almendral2007dynamical}.

We want to remark that the conventional order parameter $r$ only takes into account the phase information of the nodes, and thus the information of the network topology is implicitly contained in the steady-state values of the phases [see Fig.~\ref{fig:scheme} (b)].
In contrast, the total local order parameter, being represented as the sum of the differences in pairs of phases over all links in the network [see Fig.~\ref{fig:scheme} (c)], contains the topological information directly.
Later in the paper, we show that, interestingly, the $\tilde{r}$-based results outperform $r$-based results in many cases.

\section{Derivation}\label{sec:derivation}

Our first task is to express the order parameters as a function of $\omega_0$. For this purpose, we have to find the relation between $\theta_i (t)$ and $\omega_i$. By plugging the stationary state condition $\theta_i(t) = \theta_i^\ast + \bar{\omega} t$ into the equation of motion in Eq.~\eqref{eq:eq_of_motion}, we obtain the relation
\begin{equation}
    \bar{\omega} = \omega_i - K \sum_{j=0}^N a_{ij} \sin \left ( \theta_i^\ast - \theta_j^\ast \right)
\end{equation}
with the average frequency $\bar{\omega} = \sum_{i=0}^N \omega_i/(N+1)$.
By substituting the solution for $\theta_i^\ast$ into Eqs.~\eqref{def:order_para} and \eqref{def:local_order_para}, one may find the desired forms.
Since it is impossible to solve this relation, we take the strong coupling limit.
Without loss of generality, we set our reference frame such that $\sum_{i=0}^N \theta^\ast = 0$, which leads to $|\theta_i^\ast|\ll1$ in the strong coupling limit.
The relation is then linearized as
\begin{equation}\label{eq:linear_eq_of_motion}
    K \sum_{j=0}^N L_{ij} \theta_j(t) = \omega_i - \bar{\omega}~,
\end{equation}
where $L_{ij} = k_i \delta_{ij} - a_{ij}$ represent the elements of the Laplacian matrix $\mathbf{L}$ of the network with the degree $k_i = \sum_{j=0}^N a_{ij}$ of node $i$ and Kronecker delta $\delta_{ij}$.

This system of linear equations is undetermined because $\mathbf{L}$ has a null eigenvalue.
Thus, we have to utilize a generalized inverse, called the Moore--Penrose inverse $\mathbf{L}^\dagger = \sum_{\nu = 1}^N \lambda_\nu^{-1} \boldsymbol{v}_\nu \boldsymbol{v}_\nu^{\rm T}$,
where the superscript ${\rm T}$ stands for the transpose and $\lambda_\nu$ and $\boldsymbol{v}_\nu$ are the eigenvalues in ascending order and the corresponding eigenvectors, respectively, such that $\mathbf{L} \boldsymbol{v}_\nu = \lambda_\nu \boldsymbol{v}_\nu$ for $\nu = 0, 1, ..., N$.
By denoting the vector of the natural frequencies and the steady-state phases by $\boldsymbol{\omega} = (\omega_0, \omega_1, ..., \omega_N)^{\rm T}$ and $\boldsymbol{\theta}^\ast = (\theta_0^\ast, \theta_1^\ast, ..., \theta_N^\ast)^{\rm T}$, we obtain
\begin{equation}\label{theta_ss}
    \boldsymbol{\theta}^\ast = \frac{1}{K} \mathbf{L}^\dagger
    ( \boldsymbol{\omega}
    - \bar{\omega} \boldsymbol{1})~.
\end{equation}

In the strong coupling regime,
the order parameters up to the leading order are given by
\begin{equation}
    r =
    1 -
    \frac{1}{2 (N+1)} (\boldsymbol{\theta}^\ast)^{\rm T} \boldsymbol{\theta}^\ast
    ~\textrm{and}~
    \tilde{r} =
    \sum_j k_j -
    \frac{1}{2} (\boldsymbol{\theta}^\ast)^{\rm T} \mathbf{L} \boldsymbol{\theta}^\ast~.
\end{equation}
By substituting Eq.~\eqref{theta_ss}, we obtain
\begin{equation}
    r = 1 - \frac{1}{2 (N+1) K^2} O_2
    ~\textrm{and}~
    \tilde{r} = \sum_j k_j - \frac{1}{2 K^2} O_1
\end{equation}
with the objective functions in the same form as Refs.~\onlinecite{skardal2014optimal,Lei2022unified},
\begin{equation}
    O_p =
    (\boldsymbol{\omega} -
    \bar{\omega}
    \boldsymbol{1})^{\rm T}
    (\mathbf{L}^\dagger)^p
    (\boldsymbol{\omega} -
    \bar{\omega}
    \boldsymbol{1})~.
\end{equation}
Here, we used the relation $\mathbf{L}^\dagger \mathbf{L} \mathbf{L}^\dagger = \mathbf{L}^\dagger$.
The objective functions have been coined as the adjusted Lyapunov function~\cite{Lei2022unified} (ALF) for $p=1$ and the synchrony alignment function~\cite{skardal2014optimal} (SAF) for $p=2$.
In the strong coupling regime, the optimal synchronization is attained when the objective functions are minimized.
As the minimum point $\omega_0^\ast = {\rm argmin}_{\omega_0} [O_p (\omega_0)]$ deviates from the optimal value when the coupling strength is not sufficiently large, we regard $\omega_0^\ast$ as an approximate solution.

In order to find an available form of $\omega_0^\ast$, it is convenient to represent the matrices and vectors in block form,
\begin{equation}
    \mathbf{L} =
    \begin{pmatrix}
    k_0 &
    -\tilde{\boldsymbol{1}}^{\rm T} \tilde{\mathbf{A}}\\
    -\tilde{\mathbf{A}} \tilde{\boldsymbol{1}} &
    \tilde{\mathbf{L}} + \tilde{\mathbf{A}}
    \end{pmatrix}
    ~\textrm{and}~
    \boldsymbol{\omega} =
    \begin{pmatrix}
    \omega_0 \\
    \tilde{\boldsymbol{\omega}}
    \end{pmatrix}
\end{equation}
with $\tilde{A}_{ij}=a_{0i} \delta_{ij}$,
$\tilde{\boldsymbol{\omega}} = (\omega_1, \omega_2, ..., \omega_N)^{\rm T}$,
and the Laplacian matrix $\tilde{\mathbf{L}}$ of the prior network before the growing step.
Here, the tilde is used to indicate vectors and square matrices with the dimension reduced by $1$.

While the inverse of any invertible block matrix can be easily obtained from the block inversion formula, it is complicated to find a general formula for the Moore--Penrose inverse of any block matrix.
Thus, we utilize specific properties of $\mathbf{L}$ to find $\mathbf{L}^\dagger$:
(i) its lowest eigenvalue is $\lambda_0 = 0$, (ii) the corresponding null vector is $(1/\sqrt{N+1}) \boldsymbol{1}$, and (iii) all the other eigenvalues are positive, $\lambda_i > 0$ for $i>0$, unless the network is separated.
These properties show that we can get rid of the singularity of $\mathbf{L}$ by adding $\varepsilon \mathbf{I}$ with any positive number $\varepsilon$ and the identity matrix $\mathbf{I}$.
Thus, by subtracting the lowest eigenvalue term from $(\mathbf{L} + \varepsilon \mathbf{I})^{-1}$ and then taking the limit as $\varepsilon \rightarrow 0$, we can find the Moore--Penrose inverse
\begin{equation}\label{rel:Ldagger}
    \mathbf{L}^\dagger =
    \lim_{\varepsilon \rightarrow 0}
    \left (
    (\mathbf{L} + \varepsilon \mathbf{I})^{-1} -
    \frac{1}{\varepsilon (N+1)} \boldsymbol{1} \boldsymbol{1}^{\rm T}
    \right)~.
\end{equation}
As derived in Appendix~\ref{sec:app_A},
the formula gives
\begin{equation}\label{eq:Ldaggerp}
    (\mathbf{L}^\dagger)^p =
    \mathbf{C}^{\rm T}
    \begin{pmatrix}
    0 & \tilde{\boldsymbol{0}}^{\rm T} \\
    \tilde{\boldsymbol{0}} & \tilde{\mathbf{E}}_p
    \end{pmatrix}
    \mathbf{C}~,
\end{equation}
where the auxiliary matrices are defined by
\begin{align}
    \mathbf{C} &=
    \begin{pmatrix}
    0 & \tilde{\boldsymbol{0}}^{\rm T} \\
    \tilde{\boldsymbol{0}} & \tilde{\mathbf{I}}
    \end{pmatrix} -
    \frac{1}{N+1}
    \begin{pmatrix}
    0 & \tilde{\boldsymbol{0}}^{\rm T} \\
    \tilde{\boldsymbol{1}} &
    \tilde{\mathbf{J}}
    \end{pmatrix}~,\\
    \tilde{\mathbf{E}}_p &=
    (\tilde{\mathbf{L}}+\tilde{\mathbf{A}})^{-1}
    \left(
    \left(
    \tilde{\mathbf{I}} -
    \frac{1}{N+1} \tilde{\mathbf{J}}
    \right)
    (\tilde{\mathbf{L}} +
    \tilde{\mathbf{A}})^{-1}
    \right)^{p-1} \label{eq:E_p}
\end{align}
for any integer $p$ with a vector of zeros $\tilde{\boldsymbol{0}} = (0, 0, ..., 0)^{\rm T}$ and
a matrix of ones $J_{ij} = 1$.

This result yields the explicit form of the objective functions in terms of $\omega_i$.
For convenience, we represent the objective functions as a function of $\bar{\omega}$ rather than $\omega_0$. Then they are written as quadratic functions of $\bar{\omega}$,
\begin{equation}
    O_p (\bar{\omega}) =
    \tilde{\boldsymbol{1}}^{\rm T}
    \tilde{\mathbf{E}}_p
    \tilde{\boldsymbol{1}}
    \bar{\omega}^2 -
    2 \tilde{\boldsymbol{1}}^{\rm T}
    \tilde{\mathbf{E}}_p
    \tilde{\boldsymbol{\omega}}
    \bar{\omega} +
    \tilde{\boldsymbol{\omega}}^{\rm T}
    \tilde{\mathbf{E}}_p
    \tilde{\boldsymbol{\omega}}~.
\end{equation}
Their minimum is taken when
\begin{equation}\label{rel:main_result}
    \bar{\omega} =
    \tilde{\boldsymbol{c}}_p^{\rm T}
    \tilde{\boldsymbol{\omega}}
    ~\textrm{with}~
    \tilde{\boldsymbol{c}}_p =
    \frac{\tilde{\mathbf{E}}_p \tilde{\boldsymbol{1}}}
    {\tilde{\boldsymbol{1}}^{\rm T} \tilde{\mathbf{E}}_p\tilde{\boldsymbol{1}}}~.
\end{equation}
This optimal condition is our main result.
We note that the coefficients are normalized,
$\tilde{\boldsymbol{c}}_p^{\rm T} \tilde{\boldsymbol{1}} = 1$.
The condition shows that the optimal $\omega_0$ is simply given by
the sum of all contributions from each node, which are factorized into
the product of structure-dependent coefficients and individual natural frequencies.
This reveals that $\tilde{\boldsymbol{c}}_p$ reflects the strength and the tendency of each prior node's contribution to the optimal value.

For $p=1$, we can understand the meaning of the coefficients $\tilde{\boldsymbol{c}}_1$ by considering an auxiliary dynamics where all the oscillators are identical with the same natural frequency denoted by $f_i=f$ except for the added node having a different frequency, $f_0 = f + \Delta f$. We denote the phase of these identical oscillators by $\phi_i$ to highlight that they are not of the original system.
We remark that this setup resembles systems with an extraordinary oscillator, referred to as a pacemaker~\cite{kori2004entrainment}. In such a system, however, the interaction between the pacemaker and the other nodes is non-reciprocal.

At the steady-state, Eq.~\eqref{eq:linear_eq_of_motion} gives
\begin{equation}\label{rel:Axu_ss}
\begin{pmatrix}
f + \Delta f - \bar{f} \\
(f - \bar{f}) \tilde{\boldsymbol{1}}
\end{pmatrix}
=
\begin{pmatrix}
k_0 &
- \tilde{\boldsymbol{1}}^{\rm T} \tilde{\mathbf{A}} \\
- \tilde{\mathbf{A}} \tilde{\boldsymbol{1}} &
\tilde{\mathbf{L}} + \tilde{\mathbf{A}}
\end{pmatrix}
\begin{pmatrix}
\phi_0^* \\
\tilde{\boldsymbol{\phi}}^*
\end{pmatrix}
\end{equation}
with $\bar{f} = f + \Delta f / (N + 1)$. By using the relation $\tilde{\mathbf{A}} \tilde{\boldsymbol{1}} = ( \tilde{\mathbf{L}} + \tilde{\mathbf{A}} )\tilde{\boldsymbol{1}}$, one can recast the lower block as
\begin{equation}
\Delta \tilde{\boldsymbol{\phi}}^*
=
(f - \bar{f}) \left( \tilde{\mathbf{L}} + \tilde{\mathbf{A}} \right)^{-1}
\tilde{\boldsymbol{1}}
\propto \tilde{\boldsymbol{c}}_1~
\end{equation}
with the relative phase $\Delta \tilde{\boldsymbol{\phi}}^* = \tilde{\boldsymbol{\phi}}^* - \phi_0^* \tilde{\boldsymbol{1}}$ with respect to the distinctive oscillator’s phase.
We note that the relative phase $\phi^*_i - \phi^*_0$ is a topological node property independent of any dynamical properties.
Small $\Delta \tilde{\boldsymbol{\phi}}^*$ means that the $i$-th node is prone to synchronize with the new node, intrinsically.
In this sense, one can regard $\Delta \phi^*_i$, or equivalently the coefficients $c_{1,i}$, as a measure of the synchronization anti-affinity between node $i$ and $0$.

\section{Simple examples}\label{sec:simple_ex}

To demonstrate the validity of our result, we apply it to simple solvable examples: identical oscillators in an arbitrary network, one-to-all connection in an arbitrary network, and two cliques connected by a single link.

\subsection{Identical oscillators}

Our first example is a system with identical oscillators, $\omega_i = \omega$ for $i>0$. In this case, one can easily show that regardless of the interaction structure, the ALF and SAF take the minimum at the same value $\omega_0^\ast = \omega$ due to the normalization condition of $\tilde{\boldsymbol{c}}_p$. This result is consistent with our intuition that for identical oscillators, the optimal synchronization is attained when the added oscillator resembles the existing individuals.

\subsection{One-to-all connection}

As the second example, we consider the case where the new node is connected to all the other nodes, $a_{0i}=1$ for $i>0$.
The matrix $\tilde{\mathbf{A}}$ then becomes the identity matrix $\tilde{\mathbf{A}} = \tilde{\mathbf{I}}$.
Since $(\tilde{\mathbf{L}} + \tilde{\mathbf{I}}) \tilde{\boldsymbol{1}} = \tilde{\boldsymbol{1}}$ for any Laplacian matrix $\tilde{\mathbf{L}}$, one can show that
\begin{equation}
\tilde{\mathbf{E}}_p \tilde{\boldsymbol{1}} =
\frac{1}{(N+1)^{p-1}}
\tilde{\boldsymbol{1}}
\propto
\tilde{\boldsymbol{1}}~.
\end{equation}
Thus, the optimal frequency is simply given by the average natural frequency over other nodes, $\omega_0^\ast = \bar{\omega} \equiv \sum_{i>0} \omega_i/N$.
This implies that the average natural frequency represents a simple but effective choice when the new node is connected to most of the other nodes in the network.
Later, we show that this strategy, referred to as the global average, becomes ineffective when the new node is sparsely connected.

\subsection{Two-clique network}

\begin{figure}
\centering
\includegraphics[width = 0.5\columnwidth]{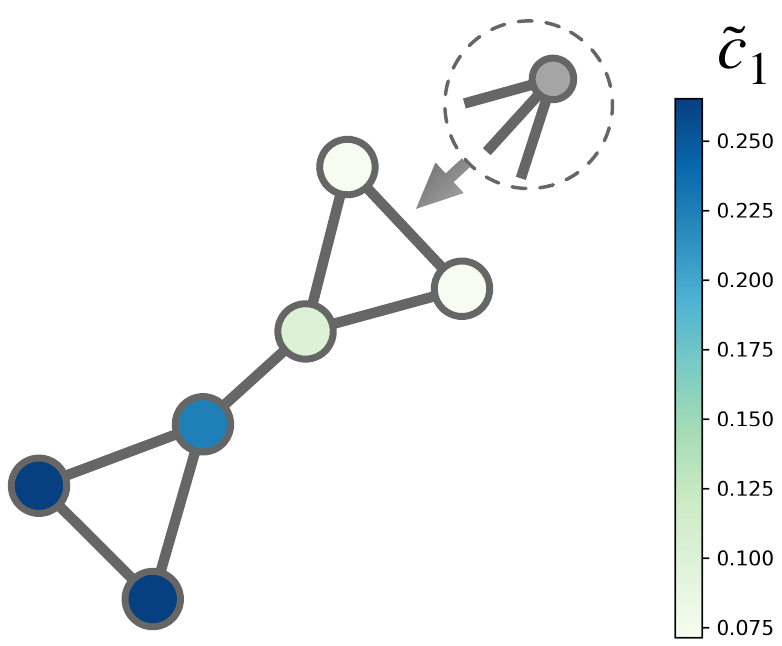}
\caption{Normalized contribution of existing nodes in the ALF method $\tilde{c}_{1}$ to determine the average frequency $\bar{\omega}$ in a two-clique model network. When a new node is introduced to one side of the model network, $\tilde{c}_{1}$ is highest at the end of the opposite side and decreases closer to the new node.
}
\label{fig:two_clique}
\end{figure}

As the last example, we consider oscillators in a network with a community structure.
As illustrated in Fig.~\ref{fig:two_clique}, we construct the base network by connecting two cliques of size $\bar{N} \equiv N/2$, denoted by $C_1$ and $C_2$, via a single link.
We call the two nodes connected to the inter-clique link `edge nodes’ and the remaining ones  `bulk nodes'.
For convenience, we assign the index $i$ in the following order:
the bulk nodes of $C_1$, the edge node of $C_1$, the edge node of $C_2$, and the bulk nodes of $C_2$.

To see the effect of inhomogeneous growth, we consider that the new node is connected to all the nodes in $C_2$.
After calculations explained in Appendix~\ref{sec:app_B},
we can obtain
\begin{equation}\label{eq:ex3_c1}
\tilde{\boldsymbol{c}}_1 \propto
\begin{pmatrix}
(\bar{N}^2 + 5\bar{N} + 2) \check{\boldsymbol{1}} \\
\bar{N}^2 + 4\bar{N} + 1 \\
3\bar{N} +1 \\
(2\bar{N} + 1) \check{\boldsymbol{1}}
\end{pmatrix}
\end{equation}
and
\begin{equation}\label{eq:ex3_c2}
\tilde{\boldsymbol{c}}_2 \propto
\begin{pmatrix}
(\bar{N}^5 + 7\bar{N}^4 + 17\bar{N}^3 + 24\bar{N}^2 + 13\bar{N} + 2) \check{\boldsymbol{1}}\\
\bar{N}^5 + 6\bar{N}^4 + 12\bar{N}^3 + 14\bar{N}^2 + 4\bar{N} -1 \\
\bar{N} (\bar{N}^3 + 4\bar{N}^2 + 10\bar{N} + 5) \\
(2\bar{N} + 1)^2 \check{\boldsymbol{1}}
\end{pmatrix}~,
\end{equation}
where the symbol $\check{\cdot}$ is used to represent vectors of dimension $\bar{N}-1$.

These results show that the coefficients are in descending order, implying that the contributions of the neighbors in $C_2$ are weaker than those of the nodes in the opposite clique $C_1$.
In other words, the addition of a node that behaves similarly to the opposing cluster helps the community cluster to synchronize with the opposite cluster.
This feature is more distinct for $\tilde{\boldsymbol{c}}_2$ and gets stronger as the system size $N$ increases.
In the large $\bar{N}$ limit, the coefficients become
\begin{equation}
\tilde{\boldsymbol{c}}_p =
\frac{1}{\bar{N}}
\begin{pmatrix}
\check{\boldsymbol{0}} \\
0 \\
1 \\
\check{\boldsymbol{1}}
\end{pmatrix} +
\mathcal{O} \left ( \frac{1}{\bar{N}^2} \right)
\end{equation}
for $p=1$ or $2$.
Accordingly, it is revealed that the SAF method is more sensitive to the network structure and that the optimal natural frequency has a negative correlation with that of its neighbors.
This negative correlation has also been reported in studies of optimal natural frequency allocation in static networks~\cite{skardal2014optimal,brede2008synchrony,buzna2009synchronization}.

\section{Numerical verification}\label{sec:num_verification}

\begin{figure*}
\includegraphics[width = 1.\columnwidth]{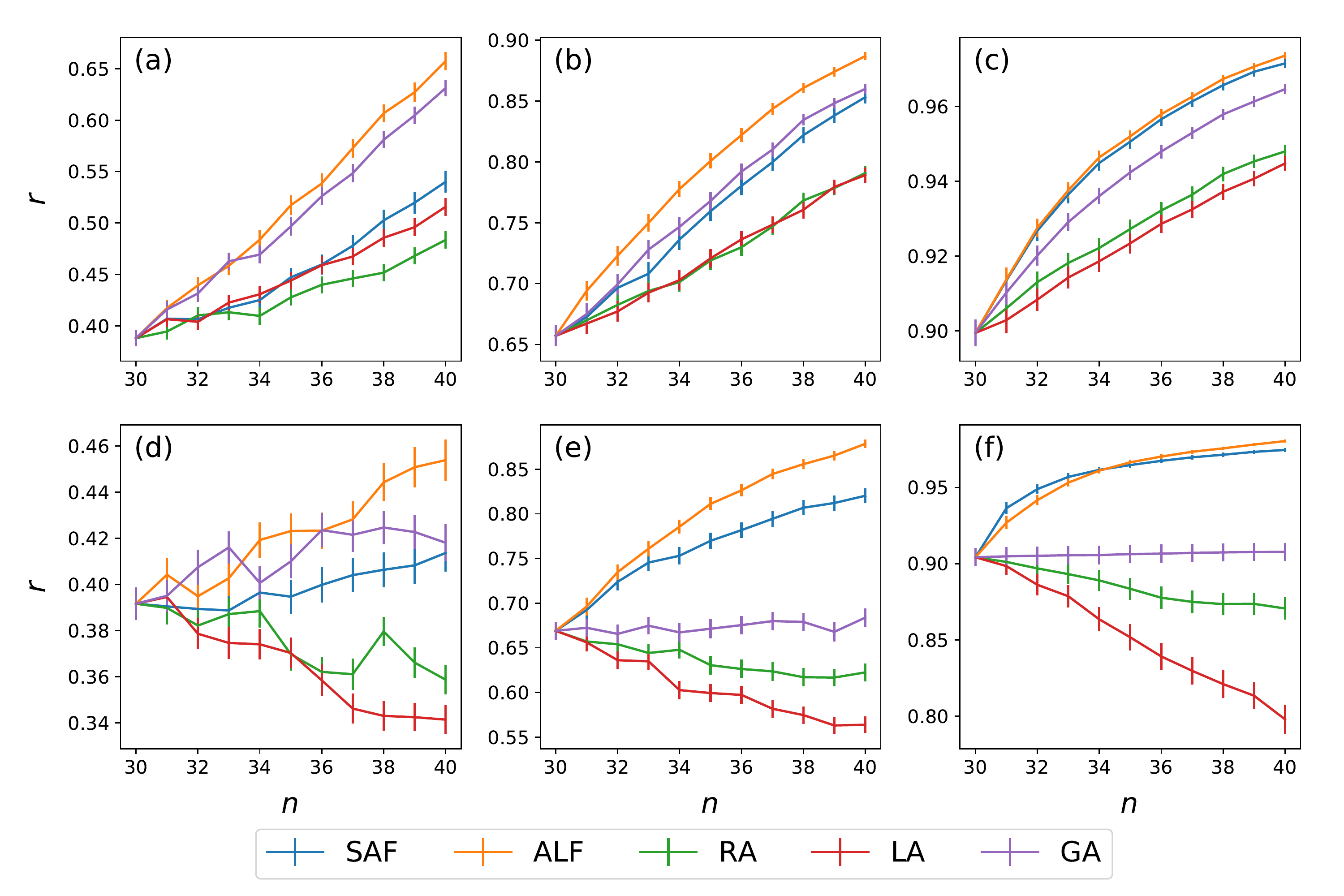}
\caption{Evolving curves of the order parameter in random growing (a,b,c) and BA (d,e,f) networks. In each panel, we use five strategies denoted in the lower legend to assign the natural frequency of the newly attached node. Each column denotes the performance of the assigning methods in weakly synchronized (a, d), moderately synchronized (b, e), and strongly synchronized (c, f) seed networks. We mark all the error bars of the order parameter with its standard error.
}
\label{fig:growing_order}
\end{figure*}

As a next step, we examine the validity of our methods for two popular growing networks: a random growing network and the Barabasi--Albert (BA) network. We construct a random growing network by connecting a new node to the others with probability $P = 0.1$. If the new node is connected to none of the existing nodes, we reject this step and try again until the new node is coupled to at least one of the existing nodes.
This constraint leads the growing random network to have different characteristics from the Erdos--Reny network as reported in Ref.~\onlinecite{callaway2001randomly}. Meanwhile, we construct a BA network by using the well-known preferential attachment algorithm~\cite{barabasi1999emergence}.

To see how the system size influences the performance of the SAF and ALF methods, we increase the system size iteratively starting from an initial seed network of size $N = 30$. We build the seed network by growing the system from a single node in accordance with the growing rule, and then assigning random numbers from the standard Gaussian distribution $\mathcal{N}(0,1)$ as the natural frequencies of the initial nodes. After that, the natural frequencies of the added nodes are taken to be the optimal $\omega_0^\ast$ obtained from Eq.~\eqref{rel:main_result}.

To demonstrate the performance of our results, we compare them to other plausible strategies: global average (GA), local average (LA), and random assignment (RA). The added node’s natural frequency is chosen to be
$\sum_{i=1}^N \omega_i/N$ for GA,
$\sum_{i=1}^N a_{0i} \omega_i/k_0$ for LA,
and a random number sampled from the same distribution as the initial natural frequency distribution for RA.
We remark that the GA method corresponds to the optimal choice obtained from the objective function $O_p$ with $p=0$.

Figure~\ref{fig:growing_order} shows the system-size dependence of the order parameter $r$ evaluated numerically in the successive growing process. Each data point is obtained from the average over 300 seed networks. We used the fourth-order Runge-–Kutta method (RK4) with a time interval of $dt = 0.001$. The relaxation time is set to be $5 \times 10^5$ steps prior to growth and $2 \times 10^5$ steps during the growing process.

The upper and lower panels in Fig.~\ref{fig:growing_order} exhibit the order parameter changes in the random growing and BA network, respectively. Each plot corresponds to different values of coupling strengths such that the initial $r$ has different values of approximately $0.4$, $0.65$, and $0.9$ (left to right). Although in principle our method may show poor performance away from the strong coupling regime, the ALF method outperforms the others even in cases with the lowest coupling strength, whereas the SAF method underperforms the GA method in the random growing network. This implies that the performance of the SAF method, as in the previous two-clique network example, is more sensitive to the interaction structure. The numerical results confirm that the ALF method provides an effective approximation in a wide range of coupling strengths.

\begin{figure*}
\centering
\includegraphics[width = 1\columnwidth]{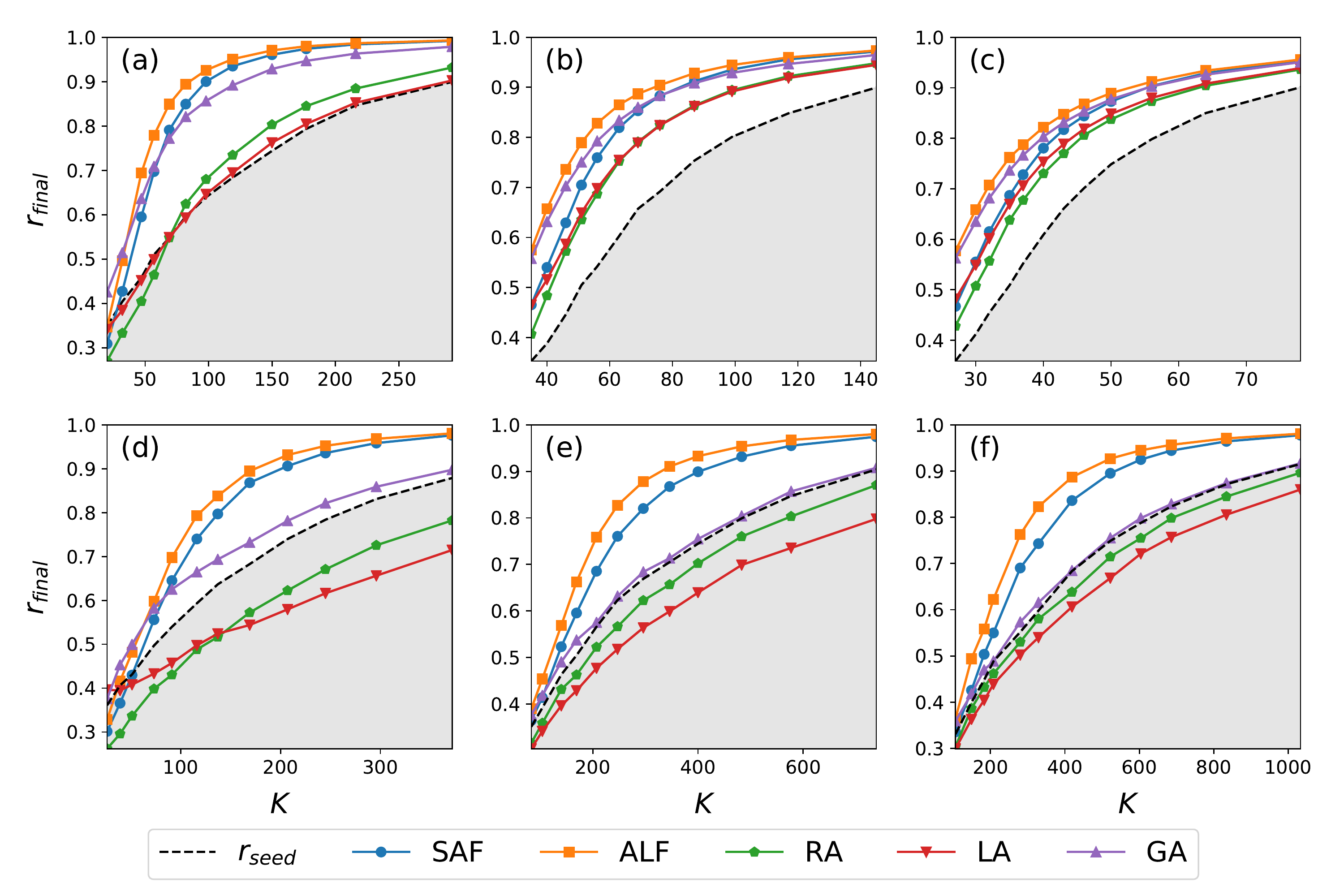}
\caption{Comparison of the initial and final state of various growing networks. The upper panels show the performance of the five assigning methods in random growing networks with (a) $N=10-20$, (b) $N=30-40$, and (c) $N=50-60$, and the lower panels exhibit the results for the BA network with the same numbers of nodes. In each graph, we plot $12$ data points with an appropriately adjusted coupling constant to uniformly represent each initial order parameter. 
}
\label{fig:seed_final}
\end{figure*}

To demonstrate the robustness of the ALF method against the coupling strength, we plot the order parameter at the terminal size as a function of the coupling strength $K$ in Fig.~\ref{fig:seed_final}.
Here, the columns represent different initial and final system sizes.
For comparison, we add black dashed curves to indicate the initial order parameter.
The left panels show that the GA method achieves the highest performance when the system size and the coupling strength are small.
But in most cases, the average-based methods are less effective, and even have a harmful influence on synchronization in the BA networks.
In contrast, the ALF and SAF methods display strong performances in both network types for different coupling constants and seed network sizes.
This certifies the effectiveness and reliability of our suggested approaches.

\section{Summary and Conclusion}\label{sec:conclusion}

In the present paper, we investigated the effect of system growth on synchronization. We have found that the addition of a new oscillator can lead to a synchronization enhancement when we assign the new oscillator with a proper natural frequency. Furthermore, we have suggested an effective and useful approximation of the optimal natural frequency by using a linearized dynamics in the strong coupling constant regime. We have applied our approximations to various examples and found that the proposed methods outperform other plausible methods based on the statistics of the existing nodes’ natural frequencies. Between the two approaches, the ALF method showed robust performance even far from the strong coupling regime.

We also observed that the performance of the ALF method is the most insensitive to the interaction structure. In particular, with stronger network heterogeneity, the other methods show relatively poorer performance. Thus, we can conclude that the ALF method is the best strategy in the sense that it leads to a notable amount of synchronization enhancement regardless of the interaction topology in a wide range of coupling strengths.

Moreover, our results revealed that the approximate optimal value is represented as the linear combination of the natural frequencies of the existing nodes. From this, we argued that the linear coefficients can be regarded as an intrinsic node property indicating a certain type of synchronization anti-affinity with respect to the added node. It would be an interesting future work to compare this quantity with the dynamical connectivity matrix~\cite{arenas2006synchronization,arenas2006synchronization2}.

It would also be an interesting future work to apply our findings to real systems described by swing equations, such as power-grid systems. In this case, we expect that it is straightforward to extend our method to second-order Kuramoto models. We believe that our work will pave the way for research into the synchronization of growing systems.

Finally, we want to remark that our approaches can also be utilized to study the synchronization of fixed-size systems.
For example, by regarding each underlying node as a newcomer, one may examine whether its natural frequency is near the optimal value.
Through this approach, we would learn how to adjust the natural frequency to enhance synchronization.

\begin{acknowledgments}
This research was supported by the KIAS individual grant No. PG074002 at the Korea Institute for Advanced Study (JMP), a National Research Foundation of Korea (NRF) grant funded by the Korean government (MSIT) No. NRF-2022R1C1C1005856 (DL and HK), and the Korea Institute of Energy Technology Evaluation and Planning (KETEP) and the Ministry of Trade, Industry \& Energy (MOTIE) of the Republic of Korea, No. 20224000000320 and No. 20224000000100 (DL and HK).
\end{acknowledgments}

\section*{Data Availability Statement}
The data that support the findings of this study are available from the corresponding author upon reasonable request.

\appendix

\section{Derivation of the main results}\label{sec:app_A}

In this appendix, we explain the details of the derivation of Eq.~\eqref{eq:Ldaggerp}.
As mentioned, the Moore--Penrose inverse $\mathbf{L}^{\dagger}$ of the Laplacian matrix can be obtained by Eq.~\eqref{rel:Ldagger}.
Since $\mathbf{L} + \varepsilon \mathbf{I}$ is invertible if $\varepsilon >0$, we can apply the block inversion formula to obtain
\begin{equation}
\left ( \mathbf{L} + \varepsilon \mathbf{I} \right )^{-1} =
\begin{pmatrix}
0 &
\tilde{\boldsymbol{0}}^{\rm T} \\
\tilde{\boldsymbol{0}} &
\tilde{\mathbf{E}}_\varepsilon
\end{pmatrix}
+ \frac{1}{k_0 + \varepsilon - \tilde{\boldsymbol{e}}^{\rm T} \tilde{\mathbf{E}}_\varepsilon^{-1} \tilde{\boldsymbol{e}}}
\begin{pmatrix}
1 &
\tilde{\boldsymbol{e}}^{\rm T} \\
 \tilde{\boldsymbol{e}} &
\tilde{\boldsymbol{e}} \tilde{\boldsymbol{e}}^{\rm T}
\end{pmatrix}
\end{equation}
with $\tilde{\mathbf{E}}_\varepsilon = ( \tilde{\mathbf{L}}+\tilde{\mathbf{A}}+\varepsilon \tilde{\mathbf{I}} )^{-1}$ and
$\tilde{\boldsymbol{e}} = \tilde{\mathbf{E}}_\varepsilon \tilde{\mathbf{A}} \tilde{\boldsymbol{1}}$.

Note that this formula is applicable only when $\tilde{\mathbf{L}} + \tilde{\mathbf{A}} + \varepsilon \tilde{\mathbf{I}}$ is invertible, which is guaranteed by the following arguments. Since the interaction network is assumed to be a connected network, the Laplacian matrix $\mathbf{L}$ must have only a single zero eigenvalue corresponding to the eigenvector $\boldsymbol{1}$.
But if $\tilde{\mathbf{L}} + \tilde{\mathbf{A}}$ is not invertible and has a zero eigenvector $\tilde{\boldsymbol{\nu}}_0$, then one can show that $(0~\tilde{\boldsymbol{\nu}}_0)^{\rm T}$ is another null vector of $\mathbf{L}$ by using $\tilde{\boldsymbol{1}}^{\rm T} \tilde{\mathbf{A}} = \tilde{\boldsymbol{1}}^{\rm T} ( \tilde{\mathbf{L}} + \tilde{\mathbf{A}} )$. Thus, from the reduction to absurdity, we prove that $\tilde{\mathbf{L}} + \tilde{\mathbf{A}}$ is invertible, so $\tilde{\mathbf{L}} + \tilde{\mathbf{A}} + \varepsilon \tilde{\mathbf{I}}$ is for a small positive $\varepsilon$.

In the small $\varepsilon$ limit, we have
\begin{align}\label{small_epsilon_ex}
\tilde{\mathbf{E}}_\varepsilon
&= \tilde{\mathbf{E}}
- \tilde{\mathbf{E}}^2 \varepsilon
+ \tilde{\mathbf{E}}^3 \varepsilon^2
+ \mathcal{O} \left ( \varepsilon^3 \right ) \\
\tilde{\boldsymbol{e}}
&= \tilde{\boldsymbol{1}}
- \tilde{\mathbf{E}} \tilde{\boldsymbol{1}} \varepsilon
+ \tilde{\mathbf{E}}^2 \tilde{\boldsymbol{1}} \varepsilon^2
+ \mathcal{O} \left ( \varepsilon^3 \right ) ~.
\end{align}
Here, we used
$\tilde{\mathbf{E}} \tilde{\mathbf{A}} \tilde{\boldsymbol{1}} = \tilde{\mathbf{E}}(\tilde{\mathbf{L}} + \tilde{\mathbf{A}})\tilde{\boldsymbol{1}} = \tilde{\boldsymbol{1}}$.
By substituting the small $\varepsilon$ expansions, we obtain
\begin{align}
&\left ( \mathbf{L} + \varepsilon \mathbf{I} \right)^{-1}
- \frac{1}{\varepsilon (N+1)} \boldsymbol{1} \boldsymbol{1}^{\rm T} \nonumber \\
&= \mathbf{E}
+ \frac{
\tilde{\boldsymbol{1}}^{\rm T} \tilde{\mathbf{E}} \tilde{\boldsymbol{1}}
}{(N+1)^2}
\mathbf{J}
- \frac{1}{N+1}
\left( \mathbf{D}^{\rm T} \mathbf{E} + \mathbf{E} \mathbf{D} \right)
+ \mathcal{O}(\varepsilon)
\end{align}
with
\begin{equation}
\mathbf{E} =
\begin{pmatrix}
0 &
\tilde{\boldsymbol{0}}^{\rm T} \\
\tilde{\boldsymbol{0}} &
\tilde{\mathbf{E}}
\end{pmatrix}
~\textrm{and}~
\mathbf{D} =
\begin{pmatrix}
0 &
\tilde{\boldsymbol{0}}^{\rm T} \\
\tilde{\boldsymbol{1}} &
\tilde{\mathbf{J}}
\end{pmatrix}~.
\end{equation}
By exploiting $(\tilde{\boldsymbol{1}}^{\rm T} \tilde{\mathbf{E}} \tilde{\boldsymbol{1}}) \mathbf{J} = \mathbf{D}^{\rm T}\mathbf{E}\mathbf{D}$ and factorizing the above equation, we can finally obtain Eq.~\eqref{eq:Ldaggerp}.

\section{Derivation of Eqs.~\eqref{eq:ex3_c1} and \eqref{eq:ex3_c2}}\label{sec:app_B}
This appendix explains how to obtain Eqs.~\eqref{eq:ex3_c1} and \eqref{eq:ex3_c2}.
They can be derived from the block inversion form of
\begin{equation}
    \tilde{\mathbf{L}} + \tilde{\mathbf{A}}
    =
    \begin{pmatrix}
    \bar{\mathbf{K}}_1 &
    - \bar{\mathbf{J}}_{\bar{N}1} \\
    - \bar{\mathbf{J}}_{1\bar{N}} &
    \bar{\mathbf{K}}_2
    \end{pmatrix}
\end{equation}
with
$\bar{\mathbf{K}}_1 = \bar{N} \bar{\mathbf{I}} - \bar{\mathbf{J}} + \bar{\mathbf{J}}_{\bar{N}\bar{N}}$,
$\bar{\mathbf{K}}_2 = (\bar{N} + 1) \bar{\mathbf{I}} - \bar{\mathbf{J}} + \bar{\mathbf{J}}_{11}$, and
$[\bar{\mathbf{J}}_{nm}]_{ij} = \delta_{ni} \delta_{mj}$.
By using a block matrix inversion formula, we can obtain
\begin{equation}\label{appB:matrix_E1}
    \left ( \tilde{\mathbf{L}} + \tilde{\mathbf{A}} \right )^{-1}
    =
    \begin{pmatrix}
    \bar{\mathbf{S}}_1^{-1} &
    \bar{\mathbf{S}}_1^{-1}
    \bar{\mathbf{J}}_{1\bar{N}}
    \bar{\mathbf{K}}_2^{-1} \\
    \bar{\mathbf{S}}_2^{-1}
    \bar{\mathbf{J}}_{\bar{N}1}
    \bar{\mathbf{K}}_1^{-1} &
    \bar{\mathbf{S}}_2^{-1}
    \end{pmatrix}
\end{equation}
with
Schur complements
$\bar{\mathbf{S}}_1 \equiv (\tilde{\mathbf{L}} + \tilde{\mathbf{A}})/\bar{\mathbf{K}}_2 = \bar{\mathbf{K}}_1 - \bar{\mathbf{J}}_{\bar{N}1} \bar{\mathbf{K}}_2^{-1} \bar{\mathbf{J}}_{1\bar{N}}$
and
$\bar{\mathbf{S}}_2 \equiv (\tilde{\mathbf{L}} + \tilde{\mathbf{A}})/\bar{\mathbf{K}}_1 = \bar{\mathbf{K}}_2 - \bar{\mathbf{J}}_{1\bar{N}} \bar{\mathbf{K}}_1^{-1} \bar{\mathbf{J}}_{\bar{N}1}$.
To find the explicit form of $\bar{\mathbf{K}}_i$,
we consider the following ansatzs,
\begin{equation}
    \bar{\mathbf{K}}_1^{-1} =
    \alpha_I \bar{\mathbf{I}}
    + \alpha_J \bar{\mathbf{J}}
    + \alpha_B \bar{\mathbf{J}}_{*\bar{N}}
    + \alpha_B \bar{\mathbf{J}}_{\bar{N}*}
    + \alpha_C \bar{\mathbf{J}}_{\bar{N}\bar{N}}
\end{equation}
and
\begin{equation}
    \bar{\mathbf{K}}_2^{-1} =
    \alpha_I \bar{\mathbf{I}}
    + \alpha_J \bar{\mathbf{J}}
    + \alpha_B \bar{\mathbf{J}}_{*1}
    + \alpha_B \bar{\mathbf{J}}_{1*}
    + \alpha_C \bar{\mathbf{J}}_{11}
\end{equation}
with $[\bar{\mathbf{J}}_{*n}]_{ij}=\delta_{nj}$,
$\bar{\mathbf{J}}_{n*} = \bar{\mathbf{J}}_{*n}^{\rm T}$,
and coefficients $\alpha_I$, $\alpha_J$, $\alpha_B$, and $\alpha_C$ to be determined.

From the condition
$\bar{\mathbf{K}}_i \bar{\mathbf{K}}_i^{-1} = \bar{\mathbf{I}}$,
we can obtain
\begin{equation}
    \alpha_I = \frac{1}{\bar{N}},~
    \alpha_J = \frac{\bar{N}+1}{\bar{N}},~
    \alpha_B = - \frac{1}{\bar{N}},~\textrm{and}~
    \alpha_C = 0
\end{equation}
for $\bar{\mathbf{K}}_1^{-1}$ and
\begin{align}
    &\alpha_I = \frac{1}{\bar{N}+1},~
    \alpha_J = \frac{\bar{N}+2}{(\bar{N}+1)(\bar{N}+3)},\nonumber\\
    ~\textrm{and}~
    &\alpha_B = \alpha_C = - \frac{1}{(\bar{N}+1)(\bar{N}+3)}
\end{align}
for $\bar{\mathbf{K}}_2^{-1}$.

Then the Schur complements are given by
\begin{align}
    \bar{\mathbf{S}}_1
    = \bar{N} \bar{\mathbf{I}}
    - \bar{\mathbf{J}}
    + \frac{\bar{N}+1}{\bar{N}+3} \bar{\mathbf{J}}_{\bar{N}\bar{N}}
    ~\textrm{and}~
    \bar{\mathbf{S}}_2
    = \left(\bar{N} + 1 \right) \bar{\mathbf{I}}
    - \bar{\mathbf{J}}.
\end{align}
To find their inverse, we again take ansatzs
\begin{equation}
    \bar{\mathbf{S}}_1^{-1} =
    \alpha_I \bar{\mathbf{I}}
    + \alpha_J \bar{\mathbf{J}}
    + \alpha_B \bar{\mathbf{J}}_{*\bar{N}}
    + \alpha_B \bar{\mathbf{J}}_{\bar{N}*}
    + \alpha_C \bar{\mathbf{J}}_{\bar{N}\bar{N}}
\end{equation}
and
\begin{equation}
    \bar{\mathbf{S}}_2^{-1} =
    \alpha_I \bar{\mathbf{I}}
    + \alpha_J \bar{\mathbf{J}}~.
\end{equation}
With the same approach, we find
\begin{equation}
    \alpha_I = \frac{1}{\bar{N}},~
    \alpha_J = \frac{\bar{N}^2+4\bar{N}+1}{\bar{N}(\bar{N}+1)},~
    \alpha_B = - \frac{1}{\bar{N}},~\textrm{and}~
    \alpha_C = 0
\end{equation}
and
\begin{equation}
    \alpha_I = \alpha_J = \frac{1}{\bar{N}+1}~.
\end{equation}
Now we can use the explicit forms of the inverse matrices to calculate the off-diagonal blocks in Eq.~\eqref{appB:matrix_E1} as
\begin{equation}
    \bar{\mathbf{S}}_1^{-1}
    \bar{\mathbf{J}}_{\bar{N}1}
    \bar{\mathbf{S}}_2^{-1}
    = \left( \bar{\mathbf{S}}_2^{-1}
    \bar{\mathbf{J}}_{1\bar{N}}
    \bar{\mathbf{S}}_1^{-1} \right)^{\rm T}
    = \frac{1}{\bar{N}+1}
    \left(
    \bar{\mathbf{J}}_{*1} + \bar{\mathbf{J}}
    \right)~,
\end{equation}
which results in
\begin{align}
    &\tilde{\mathbf{E}}_1 = 
    \left ( \tilde{\mathbf{L}} + \tilde{\mathbf{A}} \right )^{-1} \nonumber \\
    &=
    \begin{pmatrix}
    \frac{1}{\bar{N}} \bar{\mathbf{I}}
    + \frac{\bar{N}^2+4\bar{N}+1}{\bar{N}(\bar{N}+1) } \bar{\mathbf{J}}
    -\frac{1}{\bar{N}} \bar{\mathbf{J}}_{*\bar{N}}
    -\frac{1}{\bar{N}} \bar{\mathbf{J}}_{\bar{N}*} &
    \frac{1}{\bar{N}+1} \left( \bar{\mathbf{J}}_{*1} + \bar{\mathbf{J}} \right) \\
    \frac{1}{\bar{N}+1} \left( \bar{\mathbf{J}}_{1*} + \bar{\mathbf{J}} \right) &
    \frac{1}{\bar{N}+1} \left( \bar{\mathbf{I}} + \bar{\mathbf{J}} \right)
    \end{pmatrix}~.
\end{align}

By using Eq.~\eqref{eq:E_p} and multiplying it with the vector of ones, we finally obtain
\begin{equation}
\tilde{\mathbf{E}}_1 \tilde{\boldsymbol{1}} =
\frac{1}{\bar{N}+1}
\begin{pmatrix}
(\bar{N}^2 + 5\bar{N} + 2) \check{\boldsymbol{1}} \\
\bar{N}^2 + 4\bar{N} + 1 \\
3\bar{N} +1 \\
(2\bar{N} + 1) \check{\boldsymbol{1}}
\end{pmatrix}
\end{equation}
and
\begin{align}
\tilde{\mathbf{E}}_2 \tilde{\boldsymbol{1}} &=
\frac{1}{(\bar{N}+1)^2(2\bar{N}+1)} \nonumber \\
&\times \begin{pmatrix}
(\bar{N}^5 + 7\bar{N}^4 + 17\bar{N}^3 + 24\bar{N}^2 + 13\bar{N} + 2) \check{\boldsymbol{1}}\\
\bar{N}^5 + 6\bar{N}^4 + 12\bar{N}^3 + 14\bar{N}^2 + 4\bar{N} -1 \\
\bar{N} (\bar{N}^3 + 4\bar{N}^2 + 10\bar{N} + 5) \\
(2\bar{N} + 1)^2 \check{\boldsymbol{1}}
\end{pmatrix}~.
\end{align}

\nocite{*}


\begin{thebibliography}{38}%
\makeatletter
\providecommand \@ifxundefined [1]{%
 \@ifx{#1\undefined}
}%
\providecommand \@ifnum [1]{%
 \ifnum #1\expandafter \@firstoftwo
 \else \expandafter \@secondoftwo
 \fi
}%
\providecommand \@ifx [1]{%
 \ifx #1\expandafter \@firstoftwo
 \else \expandafter \@secondoftwo
 \fi
}%
\providecommand \natexlab [1]{#1}%
\providecommand \enquote  [1]{``#1''}%
\providecommand \bibnamefont  [1]{#1}%
\providecommand \bibfnamefont [1]{#1}%
\providecommand \citenamefont [1]{#1}%
\providecommand \href@noop [0]{\@secondoftwo}%
\providecommand \href [0]{\begingroup \@sanitize@url \@href}%
\providecommand \@href[1]{\@@startlink{#1}\@@href}%
\providecommand \@@href[1]{\endgroup#1\@@endlink}%
\providecommand \@sanitize@url [0]{\catcode `\\12\catcode `\$12\catcode
  `\&12\catcode `\#12\catcode `\^12\catcode `\_12\catcode `\%12\relax}%
\providecommand \@@startlink[1]{}%
\providecommand \@@endlink[0]{}%
\providecommand \url  [0]{\begingroup\@sanitize@url \@url }%
\providecommand \@url [1]{\endgroup\@href {#1}{\urlprefix }}%
\providecommand \urlprefix  [0]{URL }%
\providecommand \Eprint [0]{\href }%
\providecommand \doibase [0]{http://dx.doi.org/}%
\providecommand \selectlanguage [0]{\@gobble}%
\providecommand \bibinfo  [0]{\@secondoftwo}%
\providecommand \bibfield  [0]{\@secondoftwo}%
\providecommand \translation [1]{[#1]}%
\providecommand \BibitemOpen [0]{}%
\providecommand \bibitemStop [0]{}%
\providecommand \bibitemNoStop [0]{.\EOS\space}%
\providecommand \EOS [0]{\spacefactor3000\relax}%
\providecommand \BibitemShut  [1]{\csname bibitem#1\endcsname}%
\let\auto@bib@innerbib\@empty
\bibitem [{\citenamefont {Filatrella}, \citenamefont {Nielsen},\ and\
  \citenamefont {Pedersen}(2008)}]{power_grid}%
  \BibitemOpen
  \bibfield  {author} {\bibinfo {author} {\bibfnamefont {G.}~\bibnamefont
  {Filatrella}}, \bibinfo {author} {\bibfnamefont {A.~H.}\ \bibnamefont
  {Nielsen}}, \ and\ \bibinfo {author} {\bibfnamefont {N.~F.}\ \bibnamefont
  {Pedersen}},\ }\bibfield  {title} {\enquote {\bibinfo {title} {Analysis of a
  power grid using a kuramoto-like model},}\ }\href {\doibase
  10.1140/epjb/e2008-00098-8} {\bibfield  {journal} {\bibinfo  {journal} {The
  European Physical Journal B}\ }\textbf {\bibinfo {volume} {61}},\ \bibinfo
  {pages} {485--491} (\bibinfo {year} {2008})}\BibitemShut {NoStop}%
\bibitem [{\citenamefont {Acebr\'on}\ \emph {et~al.}(2005)\citenamefont
  {Acebr\'on}, \citenamefont {Bonilla}, \citenamefont {P\'erez~Vicente},
  \citenamefont {Ritort},\ and\ \citenamefont {Spigler}}]{circadian}%
  \BibitemOpen
  \bibfield  {author} {\bibinfo {author} {\bibfnamefont {J.~A.}\ \bibnamefont
  {Acebr\'on}}, \bibinfo {author} {\bibfnamefont {L.~L.}\ \bibnamefont
  {Bonilla}}, \bibinfo {author} {\bibfnamefont {C.~J.}\ \bibnamefont
  {P\'erez~Vicente}}, \bibinfo {author} {\bibfnamefont {F.}~\bibnamefont
  {Ritort}}, \ and\ \bibinfo {author} {\bibfnamefont {R.}~\bibnamefont
  {Spigler}},\ }\bibfield  {title} {\enquote {\bibinfo {title} {The kuramoto
  model: A simple paradigm for synchronization phenomena},}\ }\href {\doibase
  10.1103/RevModPhys.77.137} {\bibfield  {journal} {\bibinfo  {journal} {Rev.
  Mod. Phys.}\ }\textbf {\bibinfo {volume} {77}},\ \bibinfo {pages} {137--185}
  (\bibinfo {year} {2005})}\BibitemShut {NoStop}%
\bibitem [{\citenamefont {Wang}\ \emph {et~al.}(2011)\citenamefont {Wang},
  \citenamefont {Zeng}, \citenamefont {Di},\ and\ \citenamefont {Fan}}]{SBG}%
  \BibitemOpen
  \bibfield  {author} {\bibinfo {author} {\bibfnamefont {Y.}~\bibnamefont
  {Wang}}, \bibinfo {author} {\bibfnamefont {A.}~\bibnamefont {Zeng}}, \bibinfo
  {author} {\bibfnamefont {Z.}~\bibnamefont {Di}}, \ and\ \bibinfo {author}
  {\bibfnamefont {Y.}~\bibnamefont {Fan}},\ }\bibfield  {title} {\enquote
  {\bibinfo {title} {Enhancing synchronization in growing networks},}\ }\href
  {\doibase 10.1209/0295-5075/96/58007} {\bibfield  {journal} {\bibinfo
  {journal} {{EPL} (Europhysics Letters)}\ }\textbf {\bibinfo {volume} {96}},\
  \bibinfo {pages} {58007} (\bibinfo {year} {2011})}\BibitemShut {NoStop}%
\bibitem [{\citenamefont {Kuramoto}(2003)}]{kuramoto2003chemical}%
  \BibitemOpen
  \bibfield  {author} {\bibinfo {author} {\bibfnamefont {Y.}~\bibnamefont
  {Kuramoto}},\ }\href@noop {} {\emph {\bibinfo {title} {Chemical Oscillations,
  Waves, and Turbulence}}}\ (\bibinfo  {publisher} {Courier Corporation},\
  \bibinfo {year} {2003})\BibitemShut {NoStop}%
\bibitem [{\citenamefont {Acebr{\'o}n}\ \emph {et~al.}(2005)\citenamefont
  {Acebr{\'o}n}, \citenamefont {Bonilla}, \citenamefont {Vicente},
  \citenamefont {Ritort},\ and\ \citenamefont {Spigler}}]{acebron2005kuramoto}%
  \BibitemOpen
  \bibfield  {author} {\bibinfo {author} {\bibfnamefont {J.~A.}\ \bibnamefont
  {Acebr{\'o}n}}, \bibinfo {author} {\bibfnamefont {L.~L.}\ \bibnamefont
  {Bonilla}}, \bibinfo {author} {\bibfnamefont {C.~J.~P.}\ \bibnamefont
  {Vicente}}, \bibinfo {author} {\bibfnamefont {F.}~\bibnamefont {Ritort}}, \
  and\ \bibinfo {author} {\bibfnamefont {R.}~\bibnamefont {Spigler}},\
  }\bibfield  {title} {\enquote {\bibinfo {title} {The kuramoto model: A simple
  paradigm for synchronization phenomena},}\ }\href@noop {} {\bibfield
  {journal} {\bibinfo  {journal} {Reviews of modern physics}\ }\textbf
  {\bibinfo {volume} {77}},\ \bibinfo {pages} {137} (\bibinfo {year}
  {2005})}\BibitemShut {NoStop}%
\bibitem [{\citenamefont {Huang}\ \emph {et~al.}(2006)\citenamefont {Huang},
  \citenamefont {Park}, \citenamefont {Lai}, \citenamefont {Yang},\ and\
  \citenamefont {Yang}}]{huang2006abnormal}%
  \BibitemOpen
  \bibfield  {author} {\bibinfo {author} {\bibfnamefont {L.}~\bibnamefont
  {Huang}}, \bibinfo {author} {\bibfnamefont {K.}~\bibnamefont {Park}},
  \bibinfo {author} {\bibfnamefont {Y.-C.}\ \bibnamefont {Lai}}, \bibinfo
  {author} {\bibfnamefont {L.}~\bibnamefont {Yang}}, \ and\ \bibinfo {author}
  {\bibfnamefont {K.}~\bibnamefont {Yang}},\ }\bibfield  {title} {\enquote
  {\bibinfo {title} {Abnormal synchronization in complex clustered networks},}\
  }\href@noop {} {\bibfield  {journal} {\bibinfo  {journal} {Phys. Rev. Lett.}\
  }\textbf {\bibinfo {volume} {97}},\ \bibinfo {pages} {164101} (\bibinfo
  {year} {2006})}\BibitemShut {NoStop}%
\bibitem [{\citenamefont {G{\'o}mez-Gardenes}, \citenamefont {Moreno},\ and\
  \citenamefont {Arenas}(2007)}]{gomez2007paths}%
  \BibitemOpen
  \bibfield  {author} {\bibinfo {author} {\bibfnamefont {J.}~\bibnamefont
  {G{\'o}mez-Gardenes}}, \bibinfo {author} {\bibfnamefont {Y.}~\bibnamefont
  {Moreno}}, \ and\ \bibinfo {author} {\bibfnamefont {A.}~\bibnamefont
  {Arenas}},\ }\bibfield  {title} {\enquote {\bibinfo {title} {Paths to
  synchronization on complex networks},}\ }\href@noop {} {\bibfield  {journal}
  {\bibinfo  {journal} {Phys. Rev. Lett.}\ }\textbf {\bibinfo {volume} {98}},\
  \bibinfo {pages} {034101} (\bibinfo {year} {2007})}\BibitemShut {NoStop}%
\bibitem [{\citenamefont {G{\'o}mez-Garde{\~n}es}, \citenamefont {Moreno},\
  and\ \citenamefont {Arenas}(2007)}]{gomez2007synchronizability}%
  \BibitemOpen
  \bibfield  {author} {\bibinfo {author} {\bibfnamefont {J.}~\bibnamefont
  {G{\'o}mez-Garde{\~n}es}}, \bibinfo {author} {\bibfnamefont {Y.}~\bibnamefont
  {Moreno}}, \ and\ \bibinfo {author} {\bibfnamefont {A.}~\bibnamefont
  {Arenas}},\ }\bibfield  {title} {\enquote {\bibinfo {title}
  {Synchronizability determined by coupling strengths and topology on complex
  networks},}\ }\href@noop {} {\bibfield  {journal} {\bibinfo  {journal} {Phys.
  Rev. E}\ }\textbf {\bibinfo {volume} {75}},\ \bibinfo {pages} {066106}
  (\bibinfo {year} {2007})}\BibitemShut {NoStop}%
\bibitem [{\citenamefont {Arenas}\ \emph {et~al.}(2008)\citenamefont {Arenas},
  \citenamefont {D{\'\i}az-Guilera}, \citenamefont {Kurths}, \citenamefont
  {Moreno},\ and\ \citenamefont {Zhou}}]{arenas2008synchronization}%
  \BibitemOpen
  \bibfield  {author} {\bibinfo {author} {\bibfnamefont {A.}~\bibnamefont
  {Arenas}}, \bibinfo {author} {\bibfnamefont {A.}~\bibnamefont
  {D{\'\i}az-Guilera}}, \bibinfo {author} {\bibfnamefont {J.}~\bibnamefont
  {Kurths}}, \bibinfo {author} {\bibfnamefont {Y.}~\bibnamefont {Moreno}}, \
  and\ \bibinfo {author} {\bibfnamefont {C.}~\bibnamefont {Zhou}},\ }\bibfield
  {title} {\enquote {\bibinfo {title} {Synchronization in complex networks},}\
  }\href@noop {} {\bibfield  {journal} {\bibinfo  {journal} {Physics reports}\
  }\textbf {\bibinfo {volume} {469}},\ \bibinfo {pages} {93--153} (\bibinfo
  {year} {2008})}\BibitemShut {NoStop}%
\bibitem [{\citenamefont {G{\'o}mez-Gardenes}\ \emph
  {et~al.}(2011)\citenamefont {G{\'o}mez-Gardenes}, \citenamefont {G{\'o}mez},
  \citenamefont {Arenas},\ and\ \citenamefont {Moreno}}]{gomez2011explosive}%
  \BibitemOpen
  \bibfield  {author} {\bibinfo {author} {\bibfnamefont {J.}~\bibnamefont
  {G{\'o}mez-Gardenes}}, \bibinfo {author} {\bibfnamefont {S.}~\bibnamefont
  {G{\'o}mez}}, \bibinfo {author} {\bibfnamefont {A.}~\bibnamefont {Arenas}}, \
  and\ \bibinfo {author} {\bibfnamefont {Y.}~\bibnamefont {Moreno}},\
  }\bibfield  {title} {\enquote {\bibinfo {title} {Explosive synchronization
  transitions in scale-free networks},}\ }\href@noop {} {\bibfield  {journal}
  {\bibinfo  {journal} {Phys. Rev. Lett.}\ }\textbf {\bibinfo {volume} {106}},\
  \bibinfo {pages} {128701} (\bibinfo {year} {2011})}\BibitemShut {NoStop}%
\bibitem [{\citenamefont {Skardal}\ \emph {et~al.}(2013)\citenamefont
  {Skardal}, \citenamefont {Sun}, \citenamefont {Taylor},\ and\ \citenamefont
  {Restrepo}}]{skardal2013effects}%
  \BibitemOpen
  \bibfield  {author} {\bibinfo {author} {\bibfnamefont {P.~S.}\ \bibnamefont
  {Skardal}}, \bibinfo {author} {\bibfnamefont {J.}~\bibnamefont {Sun}},
  \bibinfo {author} {\bibfnamefont {D.}~\bibnamefont {Taylor}}, \ and\ \bibinfo
  {author} {\bibfnamefont {J.~G.}\ \bibnamefont {Restrepo}},\ }\bibfield
  {title} {\enquote {\bibinfo {title} {Effects of degree-frequency correlations
  on network synchronization: Universality and full phase-locking},}\
  }\href@noop {} {\bibfield  {journal} {\bibinfo  {journal} {Europhys. Lett.}\
  }\textbf {\bibinfo {volume} {101}},\ \bibinfo {pages} {20001} (\bibinfo
  {year} {2013})}\BibitemShut {NoStop}%
\bibitem [{\citenamefont {Niebur}\ \emph {et~al.}(1991)\citenamefont {Niebur},
  \citenamefont {Schuster}, \citenamefont {Kammen},\ and\ \citenamefont
  {Koch}}]{niebur1991oscillator}%
  \BibitemOpen
  \bibfield  {author} {\bibinfo {author} {\bibfnamefont {E.}~\bibnamefont
  {Niebur}}, \bibinfo {author} {\bibfnamefont {H.~G.}\ \bibnamefont
  {Schuster}}, \bibinfo {author} {\bibfnamefont {D.~M.}\ \bibnamefont
  {Kammen}}, \ and\ \bibinfo {author} {\bibfnamefont {C.}~\bibnamefont
  {Koch}},\ }\bibfield  {title} {\enquote {\bibinfo {title} {Oscillator-phase
  coupling for different two-dimensional network connectivities},}\ }\href@noop
  {} {\bibfield  {journal} {\bibinfo  {journal} {Phys. Rev. A}\ }\textbf
  {\bibinfo {volume} {44}},\ \bibinfo {pages} {6895} (\bibinfo {year}
  {1991})}\BibitemShut {NoStop}%
\bibitem [{\citenamefont {Hong}, \citenamefont {Choi},\ and\ \citenamefont
  {Kim}(2002)}]{hong2002synchronization}%
  \BibitemOpen
  \bibfield  {author} {\bibinfo {author} {\bibfnamefont {H.}~\bibnamefont
  {Hong}}, \bibinfo {author} {\bibfnamefont {M.-Y.}\ \bibnamefont {Choi}}, \
  and\ \bibinfo {author} {\bibfnamefont {B.~J.}\ \bibnamefont {Kim}},\
  }\bibfield  {title} {\enquote {\bibinfo {title} {Synchronization on
  small-world networks},}\ }\href@noop {} {\bibfield  {journal} {\bibinfo
  {journal} {Phys. Rev. E}\ }\textbf {\bibinfo {volume} {65}},\ \bibinfo
  {pages} {026139} (\bibinfo {year} {2002})}\BibitemShut {NoStop}%
\bibitem [{\citenamefont {Barahona}\ and\ \citenamefont
  {Pecora}(2002)}]{barahona2002synchronization}%
  \BibitemOpen
  \bibfield  {author} {\bibinfo {author} {\bibfnamefont {M.}~\bibnamefont
  {Barahona}}\ and\ \bibinfo {author} {\bibfnamefont {L.~M.}\ \bibnamefont
  {Pecora}},\ }\bibfield  {title} {\enquote {\bibinfo {title} {Synchronization
  in small-world systems},}\ }\href@noop {} {\bibfield  {journal} {\bibinfo
  {journal} {Phys. Rev. Lett.}\ }\textbf {\bibinfo {volume} {89}},\ \bibinfo
  {pages} {054101} (\bibinfo {year} {2002})}\BibitemShut {NoStop}%
\bibitem [{\citenamefont {Moreno}\ and\ \citenamefont
  {Pacheco}(2004)}]{moreno2004synchronization}%
  \BibitemOpen
  \bibfield  {author} {\bibinfo {author} {\bibfnamefont {Y.}~\bibnamefont
  {Moreno}}\ and\ \bibinfo {author} {\bibfnamefont {A.~F.}\ \bibnamefont
  {Pacheco}},\ }\bibfield  {title} {\enquote {\bibinfo {title} {Synchronization
  of kuramoto oscillators in scale-free networks},}\ }\href@noop {} {\bibfield
  {journal} {\bibinfo  {journal} {Europhys. Lett.}\ }\textbf {\bibinfo {volume}
  {68}},\ \bibinfo {pages} {603} (\bibinfo {year} {2004})}\BibitemShut
  {NoStop}%
\bibitem [{\citenamefont {McGraw}\ and\ \citenamefont
  {Menzinger}(2007)}]{mcgraw2007analysis}%
  \BibitemOpen
  \bibfield  {author} {\bibinfo {author} {\bibfnamefont {P.~N.}\ \bibnamefont
  {McGraw}}\ and\ \bibinfo {author} {\bibfnamefont {M.}~\bibnamefont
  {Menzinger}},\ }\bibfield  {title} {\enquote {\bibinfo {title} {Analysis of
  nonlinear synchronization dynamics of oscillator networks by laplacian
  spectral methods},}\ }\href@noop {} {\bibfield  {journal} {\bibinfo
  {journal} {Phys. Rev. E}\ }\textbf {\bibinfo {volume} {75}},\ \bibinfo
  {pages} {027104} (\bibinfo {year} {2007})}\BibitemShut {NoStop}%
\bibitem [{\citenamefont {Hong}, \citenamefont {Park},\ and\ \citenamefont
  {Tang}(2007)}]{hong2007finite}%
  \BibitemOpen
  \bibfield  {author} {\bibinfo {author} {\bibfnamefont {H.}~\bibnamefont
  {Hong}}, \bibinfo {author} {\bibfnamefont {H.}~\bibnamefont {Park}}, \ and\
  \bibinfo {author} {\bibfnamefont {L.-H.}\ \bibnamefont {Tang}},\ }\bibfield
  {title} {\enquote {\bibinfo {title} {Finite-size scaling of synchronized
  oscillation on complex networks},}\ }\href@noop {} {\bibfield  {journal}
  {\bibinfo  {journal} {Phys. Rev. E}\ }\textbf {\bibinfo {volume} {76}},\
  \bibinfo {pages} {066104} (\bibinfo {year} {2007})}\BibitemShut {NoStop}%
\bibitem [{\citenamefont {Strogatz}\ and\ \citenamefont
  {Mirollo}(1988)}]{strogatz1988collective}%
  \BibitemOpen
  \bibfield  {author} {\bibinfo {author} {\bibfnamefont {S.~H.}\ \bibnamefont
  {Strogatz}}\ and\ \bibinfo {author} {\bibfnamefont {R.~E.}\ \bibnamefont
  {Mirollo}},\ }\bibfield  {title} {\enquote {\bibinfo {title} {Collective
  synchronisation in lattices of nonlinear oscillators with randomness},}\
  }\href@noop {} {\bibfield  {journal} {\bibinfo  {journal} {J. Phys. A: Math.
  Theor.}\ }\textbf {\bibinfo {volume} {21}},\ \bibinfo {pages} {L699}
  (\bibinfo {year} {1988})}\BibitemShut {NoStop}%
\bibitem [{\citenamefont {Hong}, \citenamefont {Park},\ and\ \citenamefont
  {Choi}(2004)}]{hong2004collective}%
  \BibitemOpen
  \bibfield  {author} {\bibinfo {author} {\bibfnamefont {H.}~\bibnamefont
  {Hong}}, \bibinfo {author} {\bibfnamefont {H.}~\bibnamefont {Park}}, \ and\
  \bibinfo {author} {\bibfnamefont {M.~Y.}\ \bibnamefont {Choi}},\ }\bibfield
  {title} {\enquote {\bibinfo {title} {Collective phase synchronization in
  locally coupled limit-cycle oscillators},}\ }\href@noop {} {\bibfield
  {journal} {\bibinfo  {journal} {Phys. Rev. E}\ }\textbf {\bibinfo {volume}
  {70}},\ \bibinfo {pages} {045204} (\bibinfo {year} {2004})}\BibitemShut
  {NoStop}%
\bibitem [{\citenamefont {Hong}, \citenamefont {Park},\ and\ \citenamefont
  {Choi}(2005)}]{hong2005collective}%
  \BibitemOpen
  \bibfield  {author} {\bibinfo {author} {\bibfnamefont {H.}~\bibnamefont
  {Hong}}, \bibinfo {author} {\bibfnamefont {H.}~\bibnamefont {Park}}, \ and\
  \bibinfo {author} {\bibfnamefont {M.}~\bibnamefont {Choi}},\ }\bibfield
  {title} {\enquote {\bibinfo {title} {Collective synchronization in spatially
  extended systems of coupled oscillators with random frequencies},}\
  }\href@noop {} {\bibfield  {journal} {\bibinfo  {journal} {Phys. Rev. E}\
  }\textbf {\bibinfo {volume} {72}},\ \bibinfo {pages} {036217} (\bibinfo
  {year} {2005})}\BibitemShut {NoStop}%
\bibitem [{\citenamefont {Hong}\ \emph {et~al.}(2007)\citenamefont {Hong},
  \citenamefont {Chat{\'e}}, \citenamefont {Park},\ and\ \citenamefont
  {Tang}}]{hong2007entrainment}%
  \BibitemOpen
  \bibfield  {author} {\bibinfo {author} {\bibfnamefont {H.}~\bibnamefont
  {Hong}}, \bibinfo {author} {\bibfnamefont {H.}~\bibnamefont {Chat{\'e}}},
  \bibinfo {author} {\bibfnamefont {H.}~\bibnamefont {Park}}, \ and\ \bibinfo
  {author} {\bibfnamefont {L.-H.}\ \bibnamefont {Tang}},\ }\bibfield  {title}
  {\enquote {\bibinfo {title} {Entrainment transition in populations of random
  frequency oscillators},}\ }\href@noop {} {\bibfield  {journal} {\bibinfo
  {journal} {Phys. Rev. Lett.}\ }\textbf {\bibinfo {volume} {99}},\ \bibinfo
  {pages} {184101} (\bibinfo {year} {2007})}\BibitemShut {NoStop}%
\bibitem [{\citenamefont {Vega}, \citenamefont {V{\'a}zquez-Prada},\ and\
  \citenamefont {Pacheco}(2004)}]{vega2004fitness}%
  \BibitemOpen
  \bibfield  {author} {\bibinfo {author} {\bibfnamefont {Y.~M.}\ \bibnamefont
  {Vega}}, \bibinfo {author} {\bibfnamefont {M.}~\bibnamefont
  {V{\'a}zquez-Prada}}, \ and\ \bibinfo {author} {\bibfnamefont {A.~F.}\
  \bibnamefont {Pacheco}},\ }\bibfield  {title} {\enquote {\bibinfo {title}
  {Fitness for synchronization of network motifs},}\ }\href@noop {} {\bibfield
  {journal} {\bibinfo  {journal} {Physica A: Statistical Mechanics and its
  Applications}\ }\textbf {\bibinfo {volume} {343}},\ \bibinfo {pages}
  {279--287} (\bibinfo {year} {2004})}\BibitemShut {NoStop}%
\bibitem [{\citenamefont {Ichinomiya}(2004)}]{ichinomiya2004frequency}%
  \BibitemOpen
  \bibfield  {author} {\bibinfo {author} {\bibfnamefont {T.}~\bibnamefont
  {Ichinomiya}},\ }\bibfield  {title} {\enquote {\bibinfo {title} {Frequency
  synchronization in a random oscillator network},}\ }\href@noop {} {\bibfield
  {journal} {\bibinfo  {journal} {Phys. Rev. E}\ }\textbf {\bibinfo {volume}
  {70}},\ \bibinfo {pages} {026116} (\bibinfo {year} {2004})}\BibitemShut
  {NoStop}%
\bibitem [{\citenamefont {Restrepo}, \citenamefont {Ott},\ and\ \citenamefont
  {Hunt}(2005)}]{restrepo2005onset}%
  \BibitemOpen
  \bibfield  {author} {\bibinfo {author} {\bibfnamefont {J.~G.}\ \bibnamefont
  {Restrepo}}, \bibinfo {author} {\bibfnamefont {E.}~\bibnamefont {Ott}}, \
  and\ \bibinfo {author} {\bibfnamefont {B.~R.}\ \bibnamefont {Hunt}},\
  }\bibfield  {title} {\enquote {\bibinfo {title} {Onset of synchronization in
  large networks of coupled oscillators},}\ }\href@noop {} {\bibfield
  {journal} {\bibinfo  {journal} {Phys. Rev. E}\ }\textbf {\bibinfo {volume}
  {71}},\ \bibinfo {pages} {036151} (\bibinfo {year} {2005})}\BibitemShut
  {NoStop}%
\bibitem [{\citenamefont {Ichinomiya}(2005)}]{ichinomiya2005path}%
  \BibitemOpen
  \bibfield  {author} {\bibinfo {author} {\bibfnamefont {T.}~\bibnamefont
  {Ichinomiya}},\ }\bibfield  {title} {\enquote {\bibinfo {title}
  {Path-integral approach to dynamics in a sparse random network},}\
  }\href@noop {} {\bibfield  {journal} {\bibinfo  {journal} {Phys. Rev. E}\
  }\textbf {\bibinfo {volume} {72}},\ \bibinfo {pages} {016109} (\bibinfo
  {year} {2005})}\BibitemShut {NoStop}%
\bibitem [{\citenamefont {Lee}(2005)}]{lee2005synchronization}%
  \BibitemOpen
  \bibfield  {author} {\bibinfo {author} {\bibfnamefont {D.-S.}\ \bibnamefont
  {Lee}},\ }\bibfield  {title} {\enquote {\bibinfo {title} {Synchronization
  transition in scale-free networks: Clusters of synchrony},}\ }\href@noop {}
  {\bibfield  {journal} {\bibinfo  {journal} {Phys. Rev. E}\ }\textbf {\bibinfo
  {volume} {72}},\ \bibinfo {pages} {026208} (\bibinfo {year}
  {2005})}\BibitemShut {NoStop}%
\bibitem [{\citenamefont {Restrepo}, \citenamefont {Ott},\ and\ \citenamefont
  {Hunt}(2006)}]{restrepo2006emergence}%
  \BibitemOpen
  \bibfield  {author} {\bibinfo {author} {\bibfnamefont {J.~G.}\ \bibnamefont
  {Restrepo}}, \bibinfo {author} {\bibfnamefont {E.}~\bibnamefont {Ott}}, \
  and\ \bibinfo {author} {\bibfnamefont {B.~R.}\ \bibnamefont {Hunt}},\
  }\bibfield  {title} {\enquote {\bibinfo {title} {Emergence of synchronization
  in complex networks of interacting dynamical systems},}\ }\href@noop {}
  {\bibfield  {journal} {\bibinfo  {journal} {Physica D}\ }\textbf {\bibinfo
  {volume} {224}},\ \bibinfo {pages} {114--122} (\bibinfo {year}
  {2006})}\BibitemShut {NoStop}%
\bibitem [{\citenamefont {Skardal}, \citenamefont {Taylor},\ and\ \citenamefont
  {Sun}(2014)}]{skardal2014optimal}%
  \BibitemOpen
  \bibfield  {author} {\bibinfo {author} {\bibfnamefont {P.~S.}\ \bibnamefont
  {Skardal}}, \bibinfo {author} {\bibfnamefont {D.}~\bibnamefont {Taylor}}, \
  and\ \bibinfo {author} {\bibfnamefont {J.}~\bibnamefont {Sun}},\ }\bibfield
  {title} {\enquote {\bibinfo {title} {Optimal synchronization of complex
  networks},}\ }\href@noop {} {\bibfield  {journal} {\bibinfo  {journal} {Phys.
  Rev. Lett.}\ }\textbf {\bibinfo {volume} {113}},\ \bibinfo {pages} {144101}
  (\bibinfo {year} {2014})}\BibitemShut {NoStop}%
\bibitem [{\citenamefont {Pinto}\ and\ \citenamefont
  {Saa}(2015)}]{pinto2015optimal}%
  \BibitemOpen
  \bibfield  {author} {\bibinfo {author} {\bibfnamefont {R.~S.}\ \bibnamefont
  {Pinto}}\ and\ \bibinfo {author} {\bibfnamefont {A.}~\bibnamefont {Saa}},\
  }\bibfield  {title} {\enquote {\bibinfo {title} {Optimal synchronization of
  kuramoto oscillators: A dimensional reduction approach},}\ }\href@noop {}
  {\bibfield  {journal} {\bibinfo  {journal} {Phys. Rev. E}\ }\textbf {\bibinfo
  {volume} {92}},\ \bibinfo {pages} {062801} (\bibinfo {year}
  {2015})}\BibitemShut {NoStop}%
\bibitem [{\citenamefont {Lei}\ \emph {et~al.}(2022)\citenamefont {Lei},
  \citenamefont {Xu}, \citenamefont {Wang}, \citenamefont {Zou},\ and\
  \citenamefont {Jürgen}}]{Lei2022unified}%
  \BibitemOpen
  \bibfield  {author} {\bibinfo {author} {\bibfnamefont {Y.}~\bibnamefont
  {Lei}}, \bibinfo {author} {\bibfnamefont {X.-J.}\ \bibnamefont {Xu}},
  \bibinfo {author} {\bibfnamefont {X.}~\bibnamefont {Wang}}, \bibinfo {author}
  {\bibfnamefont {Y.}~\bibnamefont {Zou}}, \ and\ \bibinfo {author}
  {\bibfnamefont {K.}~\bibnamefont {Jürgen}},\ }\bibfield  {title} {\enquote
  {\bibinfo {title} {A unified lyapunov function for optimizing synchronization
  of coupled heterogeneous oscillators},}\ }\href@noop {} {\bibfield  {journal}
  {\bibinfo  {journal} {Preprint at
  https://doi.org/10.21203/rs.3.rs-1654969/v1}\ } (\bibinfo {year}
  {2022})}\BibitemShut {NoStop}%
\bibitem [{\citenamefont {Arenas}, \citenamefont {Diaz-Guilera},\ and\
  \citenamefont
  {P{\'e}rez-Vicente}(2006{\natexlab{a}})}]{arenas2006synchronization}%
  \BibitemOpen
  \bibfield  {author} {\bibinfo {author} {\bibfnamefont {A.}~\bibnamefont
  {Arenas}}, \bibinfo {author} {\bibfnamefont {A.}~\bibnamefont
  {Diaz-Guilera}}, \ and\ \bibinfo {author} {\bibfnamefont {C.~J.}\
  \bibnamefont {P{\'e}rez-Vicente}},\ }\bibfield  {title} {\enquote {\bibinfo
  {title} {Synchronization reveals topological scales in complex networks},}\
  }\href@noop {} {\bibfield  {journal} {\bibinfo  {journal} {Phys. Rev. Lett.}\
  }\textbf {\bibinfo {volume} {96}},\ \bibinfo {pages} {114102} (\bibinfo
  {year} {2006}{\natexlab{a}})}\BibitemShut {NoStop}%
\bibitem [{\citenamefont {Arenas}, \citenamefont {Diaz-Guilera},\ and\
  \citenamefont
  {P{\'e}rez-Vicente}(2006{\natexlab{b}})}]{arenas2006synchronization2}%
  \BibitemOpen
  \bibfield  {author} {\bibinfo {author} {\bibfnamefont {A.}~\bibnamefont
  {Arenas}}, \bibinfo {author} {\bibfnamefont {A.}~\bibnamefont
  {Diaz-Guilera}}, \ and\ \bibinfo {author} {\bibfnamefont {C.~J.}\
  \bibnamefont {P{\'e}rez-Vicente}},\ }\bibfield  {title} {\enquote {\bibinfo
  {title} {Synchronization processes in complex networks},}\ }\href@noop {}
  {\bibfield  {journal} {\bibinfo  {journal} {Physica D}\ }\textbf {\bibinfo
  {volume} {224}},\ \bibinfo {pages} {27--34} (\bibinfo {year}
  {2006}{\natexlab{b}})}\BibitemShut {NoStop}%
\bibitem [{\citenamefont {Almendral}\ and\ \citenamefont
  {D{\'\i}az-Guilera}(2007)}]{almendral2007dynamical}%
  \BibitemOpen
  \bibfield  {author} {\bibinfo {author} {\bibfnamefont {J.~A.}\ \bibnamefont
  {Almendral}}\ and\ \bibinfo {author} {\bibfnamefont {A.}~\bibnamefont
  {D{\'\i}az-Guilera}},\ }\bibfield  {title} {\enquote {\bibinfo {title}
  {Dynamical and spectral properties of complex networks},}\ }\href@noop {}
  {\bibfield  {journal} {\bibinfo  {journal} {New J. Phys.}\ }\textbf {\bibinfo
  {volume} {9}},\ \bibinfo {pages} {187} (\bibinfo {year} {2007})}\BibitemShut
  {NoStop}%
\bibitem [{\citenamefont {Kori}\ and\ \citenamefont
  {Mikhailov}(2004)}]{kori2004entrainment}%
  \BibitemOpen
  \bibfield  {author} {\bibinfo {author} {\bibfnamefont {H.}~\bibnamefont
  {Kori}}\ and\ \bibinfo {author} {\bibfnamefont {A.~S.}\ \bibnamefont
  {Mikhailov}},\ }\bibfield  {title} {\enquote {\bibinfo {title} {Entrainment
  of randomly coupled oscillator networks by a pacemaker},}\ }\href@noop {}
  {\bibfield  {journal} {\bibinfo  {journal} {Phys. Rev. Lett.}\ }\textbf
  {\bibinfo {volume} {93}},\ \bibinfo {pages} {254101} (\bibinfo {year}
  {2004})}\BibitemShut {NoStop}%
\bibitem [{\citenamefont {Brede}(2008)}]{brede2008synchrony}%
  \BibitemOpen
  \bibfield  {author} {\bibinfo {author} {\bibfnamefont {M.}~\bibnamefont
  {Brede}},\ }\bibfield  {title} {\enquote {\bibinfo {title}
  {Synchrony-optimized networks of non-identical kuramoto oscillators},}\
  }\href@noop {} {\bibfield  {journal} {\bibinfo  {journal} {Phys. Lett. A}\
  }\textbf {\bibinfo {volume} {372}},\ \bibinfo {pages} {2618--2622} (\bibinfo
  {year} {2008})}\BibitemShut {NoStop}%
\bibitem [{\citenamefont {Buzna}, \citenamefont {Lozano},\ and\ \citenamefont
  {D{\'\i}az-Guilera}(2009)}]{buzna2009synchronization}%
  \BibitemOpen
  \bibfield  {author} {\bibinfo {author} {\bibfnamefont {L.}~\bibnamefont
  {Buzna}}, \bibinfo {author} {\bibfnamefont {S.}~\bibnamefont {Lozano}}, \
  and\ \bibinfo {author} {\bibfnamefont {A.}~\bibnamefont
  {D{\'\i}az-Guilera}},\ }\bibfield  {title} {\enquote {\bibinfo {title}
  {Synchronization in symmetric bipolar population networks},}\ }\href@noop {}
  {\bibfield  {journal} {\bibinfo  {journal} {Phys. Rev. E}\ }\textbf {\bibinfo
  {volume} {80}},\ \bibinfo {pages} {066120} (\bibinfo {year}
  {2009})}\BibitemShut {NoStop}%
\bibitem [{\citenamefont {Callaway}\ \emph {et~al.}(2001)\citenamefont
  {Callaway}, \citenamefont {Hopcroft}, \citenamefont {Kleinberg},
  \citenamefont {Newman},\ and\ \citenamefont
  {Strogatz}}]{callaway2001randomly}%
  \BibitemOpen
  \bibfield  {author} {\bibinfo {author} {\bibfnamefont {D.~S.}\ \bibnamefont
  {Callaway}}, \bibinfo {author} {\bibfnamefont {J.~E.}\ \bibnamefont
  {Hopcroft}}, \bibinfo {author} {\bibfnamefont {J.~M.}\ \bibnamefont
  {Kleinberg}}, \bibinfo {author} {\bibfnamefont {M.~E.}\ \bibnamefont
  {Newman}}, \ and\ \bibinfo {author} {\bibfnamefont {S.~H.}\ \bibnamefont
  {Strogatz}},\ }\bibfield  {title} {\enquote {\bibinfo {title} {Are randomly
  grown graphs really random?}}\ }\href@noop {} {\bibfield  {journal} {\bibinfo
   {journal} {Phys. Rev. E}\ }\textbf {\bibinfo {volume} {64}},\ \bibinfo
  {pages} {041902} (\bibinfo {year} {2001})}\BibitemShut {NoStop}%
\bibitem [{\citenamefont {Barab{\'a}si}\ and\ \citenamefont
  {Albert}(1999)}]{barabasi1999emergence}%
  \BibitemOpen
  \bibfield  {author} {\bibinfo {author} {\bibfnamefont {A.-L.}\ \bibnamefont
  {Barab{\'a}si}}\ and\ \bibinfo {author} {\bibfnamefont {R.}~\bibnamefont
  {Albert}},\ }\bibfield  {title} {\enquote {\bibinfo {title} {Emergence of
  scaling in random networks},}\ }\href@noop {} {\bibfield  {journal} {\bibinfo
   {journal} {Science}\ }\textbf {\bibinfo {volume} {286}},\ \bibinfo {pages}
  {509--512} (\bibinfo {year} {1999})}\BibitemShut {NoStop}%
\end{thebibliography}
\end{document}